# A High Pressure Distorted α-Uranium (Pnma) Structure in Plutonium

S.K. Sikka

Office of the Principal Scientific Adviser to Government of India

Vigyan Bhawan Annexe, Maulana Azad Road, New Delhi – 110 011, INDIA

**Abstract**

Under pressure many rare earths and actinide metals transform to α-U structure or its lower symmetry distorted forms. We have reinterpreted the diffraction data of Dabos et al for Pu (reference 4) and find that a Am IV type distorted α-U structure in Pnma space group can explain this for its high pressure phase. The structures of this phase and α-Pu are both shown to have a distorted hcp topology. The upturn in the atomic volume of Pu at 0.1 MPa can also be rationalized on the basis of this proposal

----------

The ambient temperature α-phase of plutonium metal crystallizes in a monoclinic structure with 16 atoms in a cell of space group $P2_1/m$ [1]. On heating to its melting point of 913 K, it exhibits five more phases [2]. However, under high pressure, only one phase transition has been known to be reported in it in open literature. Akela et al alluded to a monoclinic to hcp transition under pressure in a publication on Th in 1988 [3]. Dabos et al [4], in a study up to 62 GPa, ruled out this hcp transition and instead showed that α – Pu transforms at 37 GPa to a hexagonal structure in space group $P6_3/m$ with Z=8. In this paper, we reinterpret the data of Dabos et al and find that the structure of this high pressure phase can be assigned to a four atom distorted α – U structure as found in Am under pressure and propose this to be its correct description.

α – U is an orthorhombic structure of uranium metal at ambient conditions with space group, Cmcm having 4 atoms in 4c Wyckoff positions (000,1/2,1/2,1/2 +/- (0,y,1/4). The value of y is ~ 0.1 [5]. Under pressure, this structure has been shown to occur, so far, in rare earths, Ce, Pr, Nd and the actinide, Pa [6-8]. A distorted form of this, in space group, Pnma with 4 atoms at 4c sites (+/-( x,1/4,z)) occurs in Am from 16 GPa [9] and an alloy of AmCm from 46 GPa [10]. Another form of distorted α – U phase in space group $P2_1,2_1,2_1$ has recently been shown to occur in Pr metal from 147 GPa [11]. α- uranium and these lower symmetry forms are regarded as having all the f-electrons in them as participating in bonding and are less compressible compared to their preceding phases.

Dabos et al did x-ray studies under pressure on α – Pu by employing energy dispersive (ESD) technique with one run in the angle dispersive mode using a position sensitive (PSD) detector having a 2θ range of only 16 degrees. They have given in their paper the data for d-spacings and intensities only at 55 GPa and 62 GPa. The comparison of the displayed diffraction patterns with those of Am-IV phase of Am given by Lindbaum et el shows a one to one correspondence in the observed powder peaks (see fig. 1) and we were able to index all the Bragg peaks at 55 GPa and 62 GPa on this structure. The fitted lattice constants at this pressure are a ≈ 4.70, b ≈ 4.49 and c ≈2.76 A$^o$. These may be compared with the values: a=5.093,b=4.679 and c=3.028 for Am IV at 17.6 GPa [9]. Table 1 displays the observed and calculated d-spacings. The agreement between the two is excellent. Similar was found for the data at 62 GPa. Since we do not know the operating parameters of the X-ray generator in the Dabos et al experiment, we could not calculate the intensities for the energy dispersive case. However, in Table 1, we give these for the angle dispersive mode. A good match could be obtained for the parameters x ≈ 0.45 and z ≈ 0.105 for the PSD data and in the ESD data strong and weak observed reflections are calculated to be so. It may be noted that the observation of the (201) peak rules out the α –U structure.

It is well known that structural transformations under pressure are predominantly diffusion less and involve small movements of atoms between the parent and the daughter structures. Fig. 2 compares the structures of α – Pu and the Pnma phase of Pu. This shows that the two structures have two layers along their b-axes, very similar to the topology of the hexagonal close packed structure. An approximate Pnma cell is also drawn in the α – Pu structure in Fig.2a (many variants of this cell are possible).It is thus very clear that only small movements will be required to bring about the desired transformations between these two structures In Am metal, Am IV forms from Am III which is akin to the high temperature γ-phase of Pu. This again consists of slightly distorted close packed hexagonal planes. Recently Bouchet et al [12] have proposed a mechanism for the transformation between a distorted γ-phase and Pu α phase. Again, Soderlind et al [13] have suggested by total energy calculations that α-Pu phase should be the high pressure structure for Am. This indicates that all these related structures are energetically very near to each other and transformations between them may not be surprising.

Fig. 3 shows the atomic volumes of some actinides at 0.1 MPa and 50 GPa. It shows that the upturn in equilibrium volume for Pu after Np disappears at high pressures. This upturn has been a subject of debate recently. None of the density functional based calculations have so far been able to explain this (see Jones et el [14] and references there in ). Jones et al have attributed this either to the thermal expansion between room temperature and the $0^oK$ temperature for which band structure calculations have been done or to that the 5f electrons are not fully delocalised in α – Pu. However, the volume estimated by Wallace [15] at zero Kelvin for α – Pu does not support the first suggestion. The data presented in Fig. 3 lends support to the latter explanation. That the 5 f electrons are not fully itinerant in α – Pu is in conformity with its lower value (42GPa) of the bulk modulus [16], compared with the measured bulk modulus of 104 GPa in α – U [8], ~ 118 GPa in Pa-II with α – U structure [7] and ~ 100 GPa for Am IV [9]. It is also consistent with the recent measurements of electronic heat capacities which show that α – Pu is a heavy fermion metal ( having a value 17 for electronic specfic heat coefficient (γ) , the largest so far observed for a pure element ) having in it strong f-f- correlations [17]. This finding renders not very accurate all the total energy calculations done so far  by traditional density functional techniques on α –Pu assuming fully delocalised 5f electrons.

The author is grateful to Dr. R. Chidambaram and Dr.S.M.Sharma for useful discussions.

**Table 1**     Observed and calculated diffraction date for Pnma structure of Pu at 55 GPa

| $d_{obs}$ (Å) | | h | k | l | $d_{cal}$ (Å) | $I_{obs}$ | | $I_{cal}$ |
|---|---|---|---|---|---|---|---|---|
| ESD | PSD | | | | | (ESD 7°) | (PSD) | |
| | 2.40 | 1 | 0 | 1 | 2.380 | | 95 | 100 |
| 2.352 | 2.36 | 0 | 1 | 1 | 2.351 | 100 | 100 | 64 |
| | | 2 | 0 | 0 | 2.350 | | | 56 |
| 2.243 | 2.25 | 0 | 2 | 0 | 2.245 | 38 | 90 | 76 |
| 2.087 | 2.10 | 1 | 1 | 1 | 2.103 | 30 | 30 | 15 |
| | | 2 | 1 | 0 | 2.082 | 44 | | 42 |
| 1.779 | 1.79 | 2 | 0 | 1 | 1.789 | 9 | 5 | 10 |
| 1.655 | | 2 | 0 | 1 | 1.662 | | | 31 |
| | 1.64 | 1 | 2 | 1 | 1.633 | 48 | 26 | 69 |
| 1.625 | | 2 | 2 | 0 | 1.623 | | | 39 |
| 1.396 | | 2 | 2 | 1 | 1.399 | 15 | | 10 |
| 1.359 | | 3 | 0 | 1 | 1.362 | | | 7 |
| 1.314 | | 0 | 3 | 1 | 1.316 | | | 11 |
| 1.304 | | 3 | 1 | 1 | 1.304 | 100 | | 24 |
| 1.270 | | 1 | 1 | 2 | 1.270 | | | 45 |
| | | 2 | 3 | 0 | 1.262 | | | 9 |
| 1.139 | | 1 | 2 | 2 | 1.140 | | | 3 |
| | | 4 | 1 | 0 | 1.137 | 34 | | 17 |
| 1.129 | | 0 | 4 | 0 | 1.122 | | | 9 |
| 1.040 | | 4 | 2 | 0 | 1.041 | 7 | | 1 |
| | | 3 | 0 | 2 | 1.035 | | | 8 |
| 1.015 | | 1 | 4 | 1 | 1.015 | | | 14 |
| | | 2 | 4 | 0 | 1.013 | 21 | | 8 |
| | | 3 | 1 | 2 | 1.010 | | | 8 |
| 1.001 | | 3 | 3 | 1 | 1.008 | | | 10 |

Figure Captions

Fig.1    A comparison of the observed diffraction pattern of Pu at 55 GPa [4] with that of Am IV at 89 GPa [9] shown as inset. Pu data is with Mo radiation while $\lambda = 0.3738$ A$^0$ for Am.

Fig.2    (a) Projection along the b-axis of the structure of $\alpha$-Pu at 0.1 MPa. (b) projection along the b-axis of the Pnma structure of Pu at 55 GPa. The atoms in black are at y=1/4 and those in white are at y= -1/4. Pnma type cells are shown by dotted lines in both the figures.

Fig.3    Atomic volumes of the actinide metals (Th-Am) at 0.1 MPa and 50 GPa.

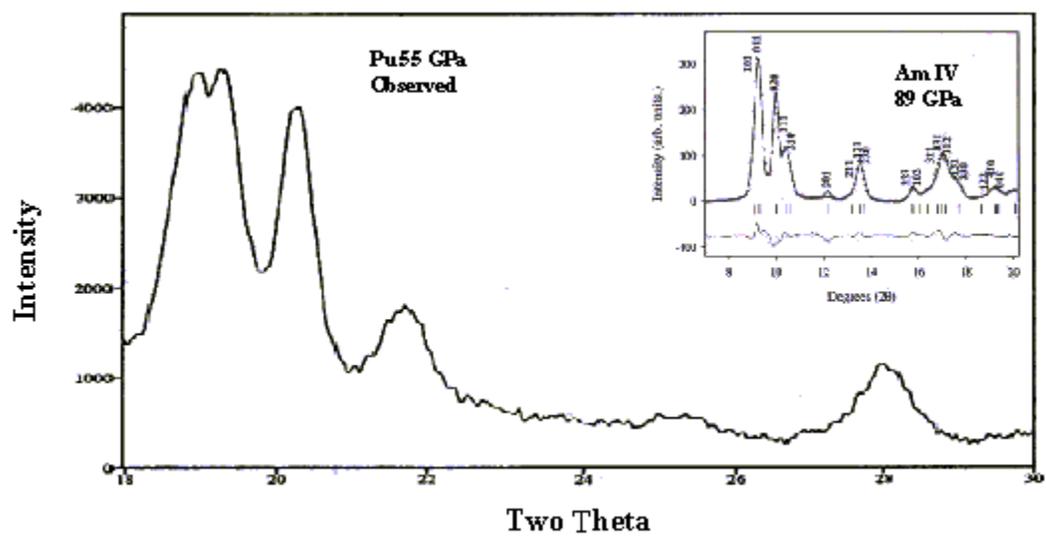

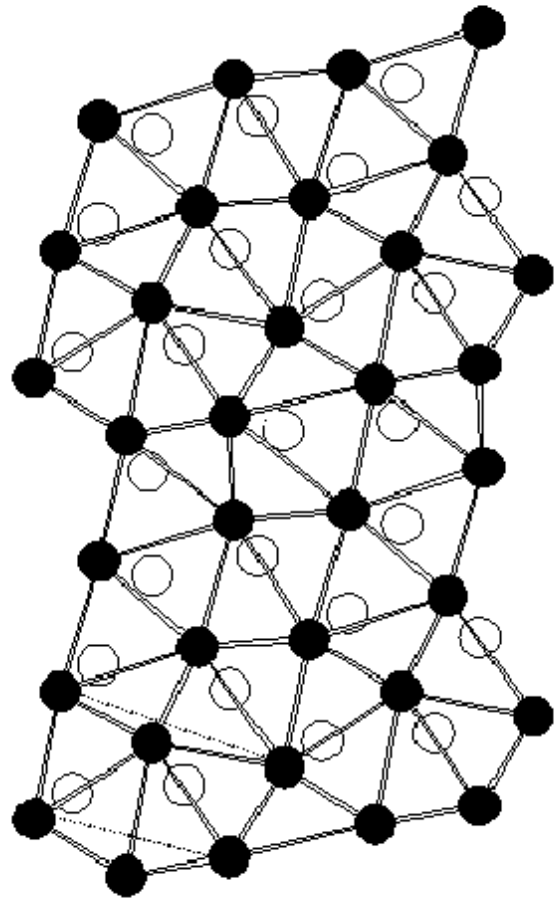

a

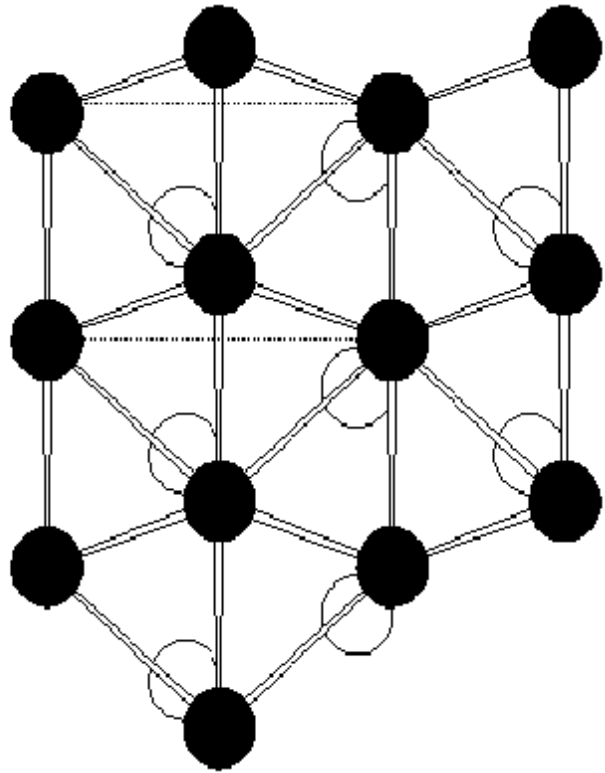

b

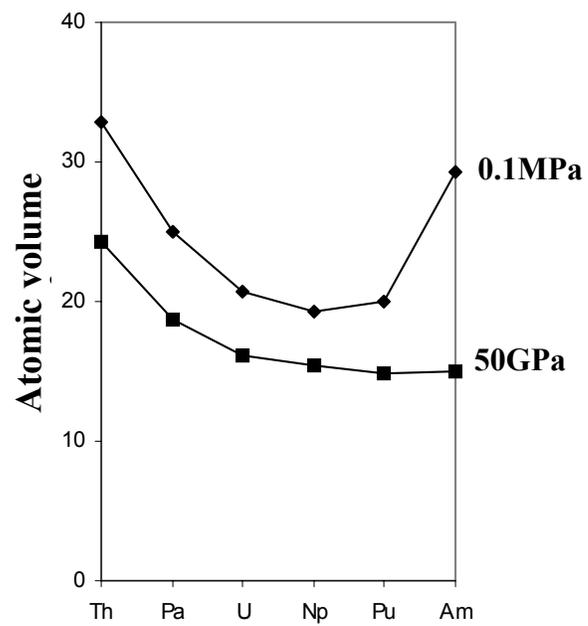